\documentclass{article}
\usepackage{amsthm, amsmath, amsfonts, amssymb}
\usepackage{charter, graphicx, mathrsfs}
\usepackage{authordate1-4}

\begin{document}

\renewcommand{\qedsymbol}{$\blacksquare$}
\newtheorem{thm}{Theorem} [section]     
\newtheorem{cor}[thm]{Corollary} 
\newtheorem{lem}[thm]{Lemma} 
\newtheorem{prop}[thm]{Proposition} 
\newtheorem{rem}[thm]{Remark}

\theoremstyle{definition}
\newtheorem{dfn}[thm]{Definition}
\newtheorem{ex}[thm]{Example}

\title{Paraconsistency and Topological Semantics}
\author{Can Ba\c{s}kent}

\date{
\begin{small}
Department of Computer Science, The Graduate Center \\ The City University of New York \\ \texttt{cbaskent@gc.cuny.edu} ~ ~ ~ \texttt{www.canbaskent.net} \\ \today
\end{small}
}

\maketitle

\section{Introduction and Motivation}

\subsection{What is Paraconsistency?}

The well-studied notion of deductive explosion describes the situation where any formula can be deduced from an inconsistent set of formulas. In other words, in deductively explosive logics, we have $\{ \varphi, \neg \varphi \} \vdash \psi$ for all formulas $\varphi, \psi$ where $\vdash$ is a logical consequence relation. In this respect, both ``classical'' and intuitionistic logics are known to be deductively explosive. Paraconsistent logic, on the other hand, is the umbrella term for logical systems where the logical consequence relation is not explosive. Variety of philosophical and logical objections can be raised against paraconsistency, and almost all of these objections can be defended in a rigorous fashion. Here, we will not be concerned about the philosophical implications of it, yet we refer the reader to the following references for comprehensive defenses of paraconsistency with a variety of well-structured applications chosen from mathematics and philosophy with a rigorous history of the subject \cite{cos,pri0,pri1,pri4}.

\subsection{Why Topologies?}

In this work, we investigate the relationship between paraconsistency and some topological spaces. As it is widely known, paraconsistency has many occurrences in mathematics spanning a wide range from model theory to set theory \cite[Chapters 2 and 3]{pri}. In this paper, we present some further applications of paraconsistency in modal logic with topological semantics.

Nevertheless, the use of topological semantics for paraconsistent logic is not new. To our knowledge, the earliest work discussing the connection between inconsistency and topology goes back to Goodman \cite{good}\footnote{Thanks to Chris Mortensen for pointing this work out. Even if the paper appeared in 1981, the work had been carried out around 1978. In his paper, Goodman indicted that the results were based on an early work that appeared in 1978 only as an abstract.}. In his paper, Goodman discussed ``pseudo-complements'' in a lattice theoretical setting and called the topological system he obtains ``anti-intuitionistic logic''. In a recent work, Priest discussed the dual of the intuitionistic negation operator and considered that operator in topological framework \cite{pri2}. Similarly, Mortensen discussed topological separation principles from a paraconsistent and paracomplete point of view and investigated the theories in such spaces \cite{mor}. Similar approaches from modal perspective was discussed by B\'{e}ziau, too \cite{bez}.

The organization of the paper is as follows. First, we will present the topological basics of our subject in a nutshell. Then, we will point out the connections between topological modal semantics and paraconsistency. Afterwards, we will make some further observations between different types of topologies and paraconsistency. Finally, we will conclude with possible research directions for future work underlining the fact that the field is rather unexplored.

\section{Basics}

\subsection{Definitions}

The history of the topological semantics of (modal) logics can be traced back to early 1920s making it the first semantics for variety of modal logics \cite{gold2}. The major revival of the topological semantics of modal logics and its connections with algebras, however, is due to McKinsey and Tarski \cite{mck,mck1}. In this section, we will briefly mention the basics of topological semantics in order to be able build our future constructions. We give two equivalent definitions of topological spaces here for our purposes.

\begin{dfn}[Topological Space] {\label{open-top}}
The structure $\langle S, \sigma \rangle$ is called a topological space if it satisfies the following conditions.
\begin{enumerate}
\item $S \in \sigma $ and $\emptyset \in \sigma $.
\item $\sigma $ is closed under arbitrary unions and under finite intersections.
\end{enumerate}
\end{dfn}

\begin{dfn}[Topological Space]{\label{close-top}}
The structure $\langle S, \tau \rangle$ is called a topological space if it satisfies the following conditions.
\begin{enumerate}
\item $S \in \tau $ and $\emptyset \in \tau$.
\item $\tau $ is closed under finite unions and under arbitrary intersections.
\end{enumerate}
\end{dfn}

Collections $\sigma$ and $\tau$ are called topologies. The elements of $\sigma$ are called \emph{open} sets whereas the elements of $\tau$ are called \emph{closed} sets. A set is called open if its complement in the same topology is a closed set and vice versa.

Functions can easily be defined on topological spaces. Recall that a function is called \emph{continuous} if the inverse image of an open (respectively, closed) set is open (respectively, closed), and a function is called \emph{open} if the image of an open (respectively, closed) set is open (respectively, closed). Moreover, two topological spaces are called \emph{homeomorphic} if there is function from one to the other which is a continuous bijection with a continuous inverse. Two continuous functions are called \emph{homotopic} if there is a continuous deformation between the two. Homotopy is then an equivalence relation and gives rise to homotopy groups which is a foundational subject in algebraic topology.

\subsection{Semantics}

In our setting, we will denote the set of propositional variables with $P$. We will use the language of propositional modal logic with the modality $\Box$, and we will define the dual $\Diamond$ in the usual sense. Therefore, we will construct the language of the basic unimodal logic recursively in the standard fashion.

In topological semantics, the modal operator for necessitation corresponds to the topological \emph{interior} operator $\mathsf{Int}$ where $\mathsf{Int}(O)$ is the largest open set contained in $O$. Furthermore, one can dually associate the topological closure operator $\mathsf{Clo}$ with the possibility modal operator $\Diamond$ where the closure $\mathsf{Clo}(O)$ of a given set $O$ is the smallest closed set that contains $O$.

Before connecting topology and modal logic, let us set a piece of notation and terminology. The extension, i.e. the points at which the formula is satisfied, of a formula $\varphi$ in the model $M$ will be denoted as $[\varphi]^{M}$. We will omit the superscript if the model we are working with is obvious. Moreover, by a \emph{theory}, we will mean a deductively closed set of formulae.

The extensions of Boolean cases are obvious. However, the extension of a modal formula $\Box \varphi$ will then be associated with an open set in the topological system. Thus, we will have $[\Box \varphi] = \mathsf{Int}([\varphi])$. Similarly, we will put $[\Diamond \varphi] = \mathsf{Clo}([\varphi])$. This means that in the basic setting, topological entities such as open or closed sets appear only with modalities. 

However, we can take one step further and suggest that extension of \emph{any} propositional variable will be an open set \cite{mor}. In that setting, conjunction and disjunction works fine for finite intersections and unions. Nevertheless, the negation can be difficult as the complement of an open set is generally not an open set, thus may not be the extension of a formula in the language. For this reason, we will need to use a new negation symbol $\dot{\sim}$ that returns the open complement (interior of the complement) of a given set. We call such systems \emph{paracomplete topological models}.

A similar idea can also be applied to closed set topologies where we stipulate that the extension of any propositional variable will be a closed set. In order to be able to avoid a similar problem with the negation, we stipulate yet another negation operator which returns the closed complement (closure of the complement) of a given set. In this setting, we will use the symbol $\sim$ that returns the closed complement of a given set. We call such systems \emph{paraconsistent topological models}.

Therefore, we generate two classes of logics with different (yet quite similar) syntax and semantics. Paracomplete topological models are generated with $\dot{\sim}, \wedge$ and $\Box$ from the set of propositional variables $P$. Paraconsistent topological models are generated with $\sim, \wedge$ and $\Box$ from the set of propositional variables $P$. Let us now make the notation clear. We will call the open set topologies which are the basis of paracomplete topological models as $\sigma$. Moreover, we will call the closed set topology which is the basis of paraconsistent topological model as $\tau$. Therefore, we make a clear cut distinction between paracomplete and paraconsistent logics and their models. Paraconsistent topological logic, in this sense, is the logic with the negation symbol $\sim$ whereas paracomplete topological logic is the logic with the negation symbol $\dot{\sim}$. The prior uses closed set topologies [extension of every proposition is closed] in its semantics, the latter uses open set topologies [extension of every proposition is open].

Now, let us consider the boundary $\partial(\cdot)$ of a set $X$ where $\partial(X)$ is defined as $\partial(X) := \mathsf{Clo}(X) - \mathsf{Int}(X)$. Consider now, for a given formuala $\varphi$, the boundary of its extension $\partial([\varphi])$ in $\tau$. Let $x \in \partial([\varphi])$. Since $[\varphi]$ is open, $x \notin [\varphi]$. Similarly, $x \notin [\dot{\sim}\varphi]$ as the open complement is also open by definition. Thus, neither $\varphi$ nor $\dot{\sim} \varphi$ is true at the boundary. Thus, in $\tau$, any theory that includes the theory of the propositions that are true at the boundary is incomplete.

Consequently, we can make a similar observation about the boundary points in $\sigma$. Now, take $x \in \partial([\varphi])$ where $[\varphi]$ is a closed set in $\sigma$. By the above definition, since we have $x \in \partial([\varphi])$, we obtain $x \in [\varphi]$ as $[\varphi]$ is closed. Yet, $\partial[(\varphi)]$ is also included in $[\sim\varphi]$ which we have defined as a closed set. Thus, by the same reasoning, we conclude $x \in [\sim \varphi]$. Thus, $x \in [\varphi \wedge \sim \varphi]$ yielding that $x \models \varphi \wedge \sim \varphi$. Therefore, in $\sigma$, any theory that includes the boundary points will be inconsistent. In this respect, a paracomplete topological model $M$ is a tuple $M = \langle S, \sigma, V \rangle$ where $\langle S, \sigma \rangle$ is an open set topology. We associate such model with a syntax that uses the negation symbol $\dot{\sim}$. Similarly, the model $M' = \langle T, \tau, V \rangle$ where $\langle T, \tau \rangle$ is a closed set topology is associated with a syntax that uses the negation symbol $\sim$, and therefore will be called paraconsistent topological model. In each cases, we call $V$ is a valuation function taking propositional variables from $P$ and returns subsets of $S$ or $T$ respectively.

So far, we have recalled how paracomplete and paraconsistent logics can be obtained in a topological setting. However, an immediate observation yields that since extensions of every formulae in $\sigma$ (respectively in $\tau$) are open (respectively, closed), the topologies which are obtained in both paraconsistent and paracomplete logics are discrete. This observation may trivialize the matter as, for instance, discrete spaces with the same cardinality are homeomorphic.

\begin{prop}
Let $M = \langle T, \tau, V \rangle$ and $M' = \langle S, \sigma, V \rangle$ be paraconsistent and paracomplete topological models respectively. If $|S| = |T|$, then there is a homeomorphism from a paraconsistent topological model to the paracomplete one, and vice versa. Moreover, $M$ amd $M'$ satisfy the same positive formula.
\end{prop}

\begin{proof}
Since $M$ and $M'$ are paraconsistent and paracomplete respectively, they have discrete topologies. Since their space have the same cardinality, $\langle S, \sigma \rangle$ and $\langle T, \tau \rangle$ are homeomorphic. Call the homeomorphism $f$. Then, $M, w \models \varphi$ if and only if $M, f(w) \models \varphi$ for negation free $\varphi$. The proof of this claim is a standard induction on the length of the formula. 

Moreover, we also observe that $M, w \models \sim{}\psi$ iff $M', f(w) \models \dot{\sim} \psi$ where $\psi$ is negation-free.
\end{proof}

\section{Topological Properties and Paraconsistency}

In this section, we investigate the relation between some basic topological properties and paraconsistency. Mostly, we will consider the closed set topology $\tau$ with its negation operator $\sim$ as it is the natural candidate for the semantics for the paraconsistent topological models.

Our work can be seen as an extension of Mortensen's earlier work \cite{mor}. Here we extend his approach to some other topological properties and discuss the behavior of such spaces under some special functions.

\subsection{Connectedness}

In the above section, we observed that boundary points play a central role in paraconsistent theories defined in topological spaces. One of the immediate topological properties that comes to mind when one deals with boundary is \emph{connectedness}. A topological space is called \emph{connected} if it is not the union of two disjoint non-empty open sets. The same definition works if we replace ``open sets'' with ``closed sets''. Formally, a set $X$ is called \emph{connected} if for two non-empty open (respectively closed) subsets $A, B$, we have $X = A \cup B$; then consequently we have $A \cap B \neq \emptyset$. Moreover, in any connected topological space, the only subsets with empty boundary are the space itself and the empty set \cite{bou}. Moreover, yet another notion in geometric topology is \emph{connected component} which is a maximal connected subspace of a given space. In this respect, we can separate topological spaces into their connected components. Also, note that connectedness is \emph{not} definable in the (classical) modal language \cite{cat}. 

Based on this definition, now establish a relation between connected spaces and theories. For this reason, we now define \emph{connected formulas} as follows.

\begin{dfn} 
A formula $\varphi$ is called connected if for any two formulae $\alpha_{1}$ and $\alpha_{2}$ with non-empty open (or dually, closed) extensions, if $\varphi \equiv \alpha_{1} \vee \alpha_{2}$, then we have $[\alpha_{1} \wedge \alpha_{2}] \neq \emptyset$. We will call a theory $T$ connected, if it is generated by a set of connected formulas. 
\end{dfn}

This definition identifies formulas with their extensions. Therefore, a connected formula $\varphi$ is actually considered as the set $[\varphi]$ at which it is true. Based on the above definition, we observe the following.

\begin{prop} Every connected formula is satisfiable in some connected (classical) topological space.
\end{prop}

\begin{proof}
Let $\varphi$ be a connected formula and $M = \langle W, \nu, V \rangle$ a (classical) topological space where for some $w \in W$, $w \models \varphi$. Then, define a connected subspace $M|\varphi = \langle W_\varphi, \nu_\varphi, V_\varphi \rangle$ as follows. Let $W_\varphi = W \cap [\varphi]^M$ so that $W_\varphi = [\varphi]^{M|\varphi}$. Notice that $W_\varphi \neq \emptyset$ as $w \in W_\varphi$. The topology $\nu_\varphi$ then is defined as follows $\nu_\varphi = \{ O \cap W_\varphi : O \in \nu \}$. It is easy to verify that $\nu_\varphi$ is indeed a topology (in fact the induced topology), so we skip it. Valuation $V$ is restricted in the usual sense. Now, we need to show that $\nu_\varphi$ is connected.

Now, take any two formulae $\alpha_{1}$ and $\alpha_{2}$ with non-empty open extensions in $M|\varphi$. Observe that if $\varphi \equiv \alpha_{1} \vee \alpha_{2}$, then $[\alpha_{1} \wedge \alpha_{2}] \neq \emptyset$. Since $W_\varphi = [\varphi]$, and the extensions $[\alpha_1]$ and $[\alpha_2]$ are nonempty by the condition, this shows that the space $W_\varphi$ is connected with respect to the topology $\nu_\varphi$.
\end{proof}

Note that the way we obtained a topological submodel is a rather standart method in modal logics. A similar theorem within the context of dynamic epistemic logic showing the completeness of that logic in topological spaces also used the same construction \cite{bas8,bas13}.

\begin{cor}
Every connected theory is satisfiable in some connected (classical) topological space.
\end{cor}

So far, we have made observations in classical topological spaces. Nevertheless, connected theories may be inconsistent or incomplete in some situations.

\begin{prop}
Every connected theory in paraconsistent topological logic is inconsistent. Moreover, every connected theory in paracomplete topological logic is incomplete.
\end{prop}

\begin{proof}
Let $T$ be a connected theory generated by a set of connected formulas $\{ \varphi_{i} \}_{i}$, so $\varphi_i \in T$ for each $i$ in a closed set topology. By the earlier corollary, $T$ is satisfiable in some connected space, say $\langle W, \sigma \rangle$.

Consider an arbitrary $\varphi_i$ from the basis of $T$. Since it is a connected formula, assume that we can write it as $\varphi_i \equiv \alpha \vee \beta$ for $[\alpha \wedge \beta] \neq \emptyset$. Let $x \in \partial[\alpha \wedge \beta] \subseteq [\varphi_i]$ as we are in a closed set topology and therefore $[\varphi]$ is closed. Thus, $T$ includes $\varphi_i$ which in turn includes the theories at $x$. By our earlier remarks, this makes $T$ inconsistent in $\sigma$. 

As a special case, in paraconsistent topological logic, observe that if $\top \in T$ where $[\top] = W$, then $T$ is inconsistent as well. Take $\top \equiv p \vee \sim p$ for some propositional variable $p$. Then, $[p \wedge{\sim}p] \neq \emptyset$.

Second part of the corollary about the incomplete theories and paracomplete models can be proved similarly.
\end{proof}

The converse direction is a bit more interesting. Do connected spaces satisfy only the connected formulas?

\begin{prop}
Let $X$ be a connected topological space of closed sets with a paraconsistent topological model on it. Then, the only subtheory that is not inconsistent is the empty theory.
\end{prop}

\begin{proof}
As we mentioned earlier, in any connected topological space, the only subsets with empty boundary are the space itself and the empty set. Thus, all other subsets will have a boundary, and their theories will be inconsistent by the earlier observations. By the earlier proposition, the space itself produce an inconsistent theory. Therefore, the only theory which is not inconsistent is the empty theory.
\end{proof}

Based on this observation, we can show a more general result.

\begin{prop}
Let $X$ be a connected topological space of closed sets. Then, for a collection of non-empty theories $T_{1}, \dots, T_{n}$ with non-empty intersection $\bigcap_i T_i$, then we conclude $\bigcup_{i} T_{i}$ is inconsistent. \end{prop}

\begin{proof}
Each theory $T_i$ will have closed set of points $X_i$ that satisfies it in the given topology. Since, $\bigcap_i T_i \neq \emptyset$, we observe $\bigcap_i X_i \neq \emptyset$. Therefore, $\bigcup_i X_i$ is connected and not equal to $X$. Thus, $\bigcup_i X_i$ has a non-empty boundary and the theories generated at the boundary points will be inconsistent.
\end{proof}

These observations hint out that boundary points play a significant role in paraconsistent topologies. A basic property of boundary gives us the following observation.

\begin{prop}
Let $X$ be an arbitrary connected topological space of closed sets. Define $\overline{X} = \{ C : C = B^{c} \text{ for some } B \text{ in } X \}$. Then, $X$ and $\overline{X}$ have the same inconsistent boundary theories.
\end{prop}

\begin{proof}
Recall that for any set $S$, we have $\partial{S} = \partial{(S^{c})}$. Therefore, the subsets in $X$ and $\overline{X}$ will have the same boundary, thus the same boundary theories. 
\end{proof}

A similar result can be shown for paracomplete theories, and we leave it to the reader.

\subsection{Continuity}

A recent research program that considers topological modal logics with continuous functions were discussed in some early works \cite{art1,kre}. In these work, they associated the modalities with continuous functions as such: $\bigcirc p = f^{-1}(p)$ where $\bigcirc$ is the temporal next time operator and $f$ is a continuous function. 

In our work, we tend to diverge from the classical modal logical approach. Our focus will rather be the connection between continuous or homeomorphic functions and modal logics with an hidden agenda of applying such approaches to paraconsistent epistemic logics in future works.

An immediate theorem, which was stated and proved in variety of different work, would also work for paraconsistent logics \cite{kre}. Now, let us take two closed set topologies $\tau$ and $\tau'$ on a given set $T$ and a homeomorphism $f:\langle T, \tau \rangle \mapsto \langle T, \tau' \rangle$. Akin to a previous theorem of Kremer and Mints, we have a simple way to associate the respective valuations between two models $M$ and $M'$ which respectively depend on $\sigma$ and $\sigma'$ so that we can have a truth preservation result. Therefore, define $V'(p)= f(V(p))$. Then, we have $M \models \varphi$ iff $M' \models \varphi$.

\begin{thm}{\label{cont-thm}}
Let $M = \langle T, \tau, V \rangle$ and $M' = \langle T, \tau', V' \rangle$ be two paraconsistent topological models (where $\tau, \tau'$ are closed set topologies) with a homeomorphism $f$ from $\langle T, \tau \rangle$ to $\langle T, \tau' \rangle$. Define $V'(p)= f(V(p))$. Then $M \models \varphi$ iff $M' \models \varphi$ for all $\varphi$.
\end{thm}

\begin{proof}
The proof is by induction on the complexity of the formulae.

Let $M, w \models p$ for some propostional variable $p$. Then, $w \in V(p)$. Since we are in a paraconsistent topological model, $V(p)$ is a closed set and since $f$ is a homeomorphism $f(V(p))$ is closed as well, and $f(w) \in f(V(p))$. Thus, $M', f(w) \models p$. Converse direction is similar and based on the fact that the inverse function is also continuous.

Negation $\sim$ is less immediate. Let $M, w \models \sim \varphi$. Therefore, $w$ is in the closure of the complement of $V(\varphi)$. So, $w \in \mathsf{Clo}((V(\varphi))^{c})$. Then, $f(w) \in f(\mathsf{Clo}(V(\varphi))^{c})$. Moreover, since $f$ is bicontinuous as $f$ is a homeomorphism, we observe that $f(w) \in \mathsf{Clo}(f((V(\varphi))^{c}))$. Then, by the induction hypothesis, $f(w) \in \mathsf{Clo}((V'(\varphi))^{c})$ yielding $M', f(w) \models \sim \varphi$. Converse direction is also similar.

We leave the conjunction case to the reader and proceed to the modal case. Assume $M, w \models \Diamond \varphi$. Thus, $w \in V(\Diamond \varphi)$. Thus, $w \in \mathsf{Clo}(V(\varphi))$. Then, $f(w) \in f(\mathsf{Clo}(V(\varphi)))$. Since $f$ is a homomorphism, we have $f(w) \in \mathsf{Clo}(f(V(\varphi)))$. By the induction hypothesis, we then deduce that $f(w) \in \mathsf{Clo}(V'(\varphi))$ which in turn yields that $f(w) \in V'(\Diamond \varphi)$. Thus, we deduce $M' , f(w) \models \Diamond \varphi$.

Converse direction is as expected and we leave it to the reader.
\end{proof}

Notice that the above theorem also works in paracomplete topological models, and we leave the details to the reader.

Assuming that $f$ is a homeomorphism may seem a bit strong. We can then seperate it into two chunks. One direction of the biconditional can be satisfied by continuity whereas the other direction is satisfied by the openness of $f$. 

\begin{cor}
 Let $M = \langle T, \tau, V \rangle$ and $M' = \langle T, \tau', V' \rangle$ be two paraconsistent topological models with a continuous $f$ from $\langle T, \tau \rangle$ to $\langle T, \tau' \rangle$. Define $V'(p)= f(V(p))$. Then $M \models \varphi$ implies $M' \models \varphi$ for all $\varphi$.
\end{cor}

\begin{cor}
 Let $M = \langle T, \tau, V \rangle$ and $M' = \langle T, \tau', V' \rangle$ be two paraconsistent topological models with an open $f$ from $\langle T, \tau \rangle$ to $\langle T, \tau' \rangle$. Define $V'(p)= f(V(p))$. Then $M' \models \varphi$ implies $M \models \varphi$ for all $\varphi$.
\end{cor}

Proofs of both corollaries depend on the fact that $\mathsf{Clo}$ operator commutes with continuous functions in one direction, and it commutes with open functions in the other direction. Furthermore, similar corollaries can be given for paracomplete frameworks as the $\mathsf{Int}$ operator also commutes in one direction under similar assumptions, and we leave it to the reader as well.

Furthermore, any topological operator that commutes with continuous, open and homeomorphic functions will reflect the same idea and preserve the truth\footnote{Thanks to Chris Mortensen for pointing this out.}. Therefore, these results can easily be generalized.

We can now take one step further to discuss homotopies in paraconsistent topological modal models. To our knowledge,  the role of homotopies as transformations between truth preserving continuous isomorphisms or bisimulations under some restrictions has not yet been discussed within the field of topological models of classical modal logic. Therefore, we believe our treatment is the first introduction of homotopies in topological semantics of modal logics. The reason why we start from paraconsistent (paracomplete) modal logics is the simple fact that the extension of each propositional letter is a closed (open) set which makes our task relatively easy and straightforward.

Recall that a \emph{homotopy} is a description of how two continuous function from a topological space to another can be deformed to each other. We can now state the formal definition.

\begin{dfn} Let $S$ and $S'$ be two topological spaces with continuous functions $f, f': S \rightarrow S'$. A homotopy between $f$ and $f'$ is a continuous function $H : S \times [0, 1] \rightarrow S'$ such that if $s \in S$, then $H(s, 0) = f(s)$ and $H(s, 1) = g(s)$ 
\end{dfn}

In other words, a homotopy between $f$ and $f'$ is a family of continuous functions $H_t: S \rightarrow S'$ such that for $t \in [0, 1]$ we have $H_0 = f$ and $H_1 = g$ and the map $t \rightarrow H_t$ is continuous from $[0, 1]$ to the space of all continuous functions from $S$ to $S'$. Notice that homotopy relation is an equivalence relation. Thus, if $f$ and $f'$ are homotopic, we denote it with $f \approx f'$. But, why do we need homotopies? We will now use homotopies to obtain a generalization of Theorem~\ref{cont-thm}.

Assume that we are given two topological spaces $\langle S, \sigma \rangle$ and $\langle S, \sigma' \rangle$ and a family of continuous functions $f_t$ for $t \in [0, 1]$. Define a model $M$ as $M = \langle S, \sigma, V \rangle$. Then, for each $f_t$ with $t \in [0, 1]$, define $M_t = \langle S, \sigma, V_t \rangle$ where $V_t = f_t(V)$. Then, by Theorem~\ref{cont-thm}, we observe that $M \models \varphi$ iff $M_t \models \varphi$. Now, what is the relation among $M_t$s? The obvious answer is that their valuation form a homotopy equivalance class. Let us now see how it works.

Define $H : S \times [0, 1] \rightarrow S'$ such that if $s \in S$, then $H(s, 0) = f_0(s)$ and $H(s, 1) = f_1(s)$. Then, $H$ is a homotopy. Therefore, given a (paraconsistent) topological modal model $M$, we generate a family of models $\{M_t\}_{t \in [0, 1]}$ whose valuations are generated by homotopic functions.

\begin{dfn}
Given a model $M = \langle S, \sigma, V \rangle$, we call the family of models $\{ M_t = \langle S, \sigma, V_t \rangle \}_{t \in [0, 1]}$ generated by homotopic functions and $M$ homotopic models. In the generation, we put $V_t = f_t(V)$. 
\end{dfn}

\begin{thm}
Homotopic paraconsistent (paracomplete) topological models satisfy the same modal formulae.
\end{thm}

\begin{proof}
See the above discussion.
\end{proof}

In the above discussions, we have focused on continuous functions and the homotopies they generate. We can also discuss homeomorphisms and their homotopies which generate homotopy equivalences between spaces. In that case, homotopic equivalent spaces can be continuously deformed to each other. This would give us, under the correct valuation, a stronger notion of bisimulation that we call \emph{continuous topo-bisimulation}. We will first start with the definition of topo-bisimulation before introducing continuous topo-bisimulation \cite{aie1}.

\begin{dfn}
Let two (classical) topological models $\langle S, \sigma, V \rangle$ and $\langle S', \sigma', V' \rangle$ be given, a topological bisimulation is a relation on $S \times S'$, and when two points $s$ from $S$ and $s'$ from $S'$ are topo-bisimular, they satisfy the following conditions.
\begin{enumerate}
 \item The points $s$ and $s'$ satisfy the same propositional variables.
 \item For $s \in O \in \sigma$, there is $O' \in \sigma'$ such that $s' \in O'$ and $\forall t' \in O'$, $\exists t \in O$ such that $t$ and $t'$ are topo-bisimular
 \item For $s' \in O' \in \sigma'$, there is $O \in \sigma$ such that $s \in O$ and $\forall t \in O$, $\exists t' \in O'$ such that $t$ and $t'$ are topo-bisimular
\end{enumerate} \end{dfn}

Now we can extend it to continuity.

\begin{dfn}
Let $M = \langle S, \sigma, V \rangle $ and $M' = \langle S', \sigma', V \rangle$ be two paraconsistent (paracomplete) topological models. We say $M, w$ and $M', w'$ are continuously topo-bisimular (denoted $M, w \rightleftharpoons M', w'$) if $M, w$ and $M', w'$ are topo-bisimular and there is a homeomorphism $f$ between $\langle S, \sigma \rangle$ and $\langle S', \sigma' \rangle$ such that $V' = f(V)$.
\end{dfn}

Note that in the above definition, we need a stronger notion of homeomorphism rather than just continuity as the bisimulation is a symmetric relation.

\begin{thm}
Continuously bisimular states satisfy the same modal formulae.
\end{thm}

\begin{proof}
The proof is an induction on the complexity of the formulas in the standard sense, and uses Theorem~\ref{cont-thm}.
\end{proof}

What about the converse? Can we have a property akin to Hennesy-Millner property so that for some topologies that satisfy exactly the same formulae, we can construct a homeomorphism in between? Clearly, answer to this question is positive if we are in finite spaces, and the construction is essentially the same as in the classical case. We refer the interested reader to a textbook treatment of classical modal logic to see how Hennesy-Millner property is treated \cite{bla1}.

Now, mathematically oriented reader might anticipate a second move towards homotopy groups and their use in modal logic. Note that homotopy groups essentially classifies the spaces with regard to their continuous deformability to each other, and it seems feasible to import such a concept to modal logics. Nevertheless, in order not to diverge our focus here, we will not pursue that path here, and leave if for a future work.

\subsection{Modal Direction}

This section of the paper will briefly review the modal approaches to the paraconsistency in order to make our work more self-contained.

One possible modal interpretation of paraconsistency focuses on the negation operator \cite{bez}. Under the usual alethic reading of $\Box$ and $\Diamond$ modalities, one can define an additional operator $\sim$ as $\neg \Box$, or equivalently $\sim \varphi \equiv \Diamond \neg \varphi$. Notice that this definition corresponds to our earlier definition of negation being the closed complement. For this interpretation, recall that $\Diamond$ operator needs to be takes as the $\mathsf{Clo}$ operator.

The Kripkean semantics of the new paraconsistent negation operator $\sim$ is as follows \cite{bez}. Let us take a modal model $M = \langle W, R, V \rangle$ where $R$ is a binary relation on the non-empty set of worlds $W$ and $V$ is valuation. Take an arbitrary state $w \in W$.
\begin{quote}
$\sim \varphi$ is false at $w$ if and only if $\varphi$ is true at every $v$ with $wRv$ 
\end{quote}

More technically, we have the following reasoning.
\begin{quote}
\begin{tabular}{lll}
$w \not\models \sim \varphi$ & iff & $w \not\models \neg \Box \varphi$ \\
& & $w \models \Box \varphi$ \\
& & $\forall v.(wRv \rightarrow v \models \varphi)$ \\
& & $w \models \varphi $ \\ 
\end{tabular} 
\end{quote}

Furthermore, as it was observed, $\sim$ modality is indeed S5, and furthermore an S5 logic can be given by taking $\sim$ as the primitive negation symbol with the intended interpretation. Nevertheless, for our current purposes, S4-character of that modality is sufficient, and we will not go into the details of such an S5 construction. We refer the interested reader to the following references for a further investigation of this subect \cite{bez0,bez}.

Moreover, it is easy to notice the similarity of modal negation we presented here and the topological negation that we used throughout his paper. Therefore, it is a nice exercise to import our topological results from topological semantics to Kripke semantics with the modal negation at hand. Therefore, one can define a modal negation in Kripke models that reflect the exact same negation that we used for paraconsistent topologies.

For this reason, we can offer a transformation from topological models to Kripke models which is similar to the standard translation between classical topological models and Kripke models \cite{aie1}. Given a topological paraconsistent model $M = \langle S, \sigma, V \rangle$, we put $sR_{\sigma}t$ when $s \in \mathsf{Clo}(t)$ to get a Kripke model $M_{\sigma} = \langle S, R_{\sigma}, V \rangle$. 
This transformation is truth preserving.

\begin{thm} Given a topological paraconsistent model $M$, if $M, w \models \varphi$ then $M_{\sigma}, w \models \varphi$ where $M_{\sigma}$ is obtained from $M$ by the transformation that $wR_{\sigma}v$ when $w \in \mathsf{Clo}(v)$.
\end{thm}

\begin{proof}
Induction on the complexity of the formulae, and the proof is a careful interplay between different negations. We will only show it for negation then. Note that we use $\sim$ for both paraconsistent Kripkean negation and paraconsistent closed set negation; nevertheless, the context will make it clear which one we mean. 

Let $M = \langle T, \tau, V \rangle$ be given. Assume $M, w \models \sim \varphi$. Since, the topological negation $\sim$ is the closure of the set theoretical complement, we observe that $M, w \models \Diamond \neg \varphi$. Therefore, for every closed set $U \in \tau$, there is a point $v \in U$ such that $M, w \models \neg \varphi$. Observe that since $v \in U$ for closed $U$, we observe that $w \in \mathsf{Clo}(v)$. Then, put $wR_{\tau}v$. So, in the model $M_{\tau} = \langle T, R_{\tau}, V \rangle$, we have $M_{\tau}, w \models \exists v (wR_{\tau}v \text{ and } M_{\tau}, v \models \neg \varphi)$. Then, by the usual semantics of modal logic, we observe $M_{\tau}, w \models \Diamond \neg \varphi$ which is nothing but $M_{\tau}, w \models \neg \Box \varphi$. Finally, by definition, we conclude $M_{\tau}, w \models \sim \varphi$.
\end{proof}

A well-known transformation from Kripke frames generate a topological space: in that case, opens are downward (or upward) closed sets (subtrees) in the Kripke model. It is also easy to prove that this transformation respects the truth of the formulae.

\begin{thm} Given a paraconsistent Kripke model $M$, if $M, w \models \varphi$ then $M_{R}, w \models \varphi$ where $M_{R}$ is obtained from $M$ by the transformation that the closed sets are downward closed subsets with respect to the accessibility relation $R$.
\end{thm}

This establishes the connection between paraconsistent topological models and paraconsistent Kripke models.

\section{Conclusion and Future Work}

In this work, we focused on the connection between topological spaces and paraconsistent logic. There are many open questions that we have left for further work. Some of them can be summarized as follows.
\begin{itemize}
\item How can we logically define homotopy and cohomotopy groups in paraconsistent or paracomplete topological modal models?
\item How would paraconsistency be affected under topological products?
\item What is the (paraconsistent) logic of regular sets?
\end{itemize}

Aforementioned questions pose yet another research program in which algebraic topological and algebraic geometrical ideas are utilized in non-classical modal logics. The interaction between truth and in such frameworks exhibits a novel line of research. Moreover, region based modal logics have presented variety of results about the logic of space \cite{pra}. Considering their use of regular sets within the framework of region based modal logics, it is not difficult to see a connection between region based modal logics and paraconsistent logics.

Furthermore, the strong algebraic connection between variety of topological models pose a very interesting approach. Considering the dual relation between intuitionistic and paraconsistent logics and their respective algebraic structures being Heyting and Brouwer algebras, their connection in the modal framework was also investigated \cite{rau}. Therefore, connection topological ideas with the existing algebraic work is yet another research direction for future work.

Yet another possible applications of such systems is epistemic logics where the knowers or agents can have inconsistent or incomplete belief basis. The intuitive connection between AGM update and paraconsistency within this framework is yet to be established. Moreover, within the domain of dynamic epistemic logic, paraconsistent announcements can be considered where agents may have inconsistent knowledge set, and yet maintain a sensible way to make deductions. We leave such stimulating discussions to future work.

\paragraph{Acknowledgement} I am grateful to Chris Mortensen and Graham Priest for their encouragement and comments.

\bibliographystyle{authordate3}  
\bibliography{/Users/can/Documents/Akademik/papers.bib} 
\end{document}